\documentclass[12pt]{jetp} 
\usepackage{amssymb,amsmath,amsfonts}
\usepackage{graphicx,epstopdf}

\English

\title{Anomalous Heat and Momentum Transport Arising from Surface Roughness in a Normal $^3$He Slab
} 

\author{Priya}{Sharma}
\affiliation{Centre for NanoScience and Engineering, Indian Institute of Science 560012, Bangalore, India}
\begin{document}
\abstract{I discuss heat and momentum transport in a mesoscopic film of $^3$He, confined by rough walls in the normal Fermi liquid  state. Inelastic binary quasiparticle scattering mediated by elastic scattering from the surface roughness gives rise to a coherent "mixed" scattering channel that drives anomalous transport over a range of temperature. I calculate the thermal conductivity and viscosity of the film in this regime and derive these in terms of the film thickness and autocorrelation function of the surface roughness, which enters the formulation as an independent input. This calculation can be useful in understanding and isolating the effects of confinement and surface roughness, especially in the context of exploring the superfluid state in the film.

}

\maketitle


\section{Introduction}
The effect of boundaries on the behaviour of physical systems is gaining in importance as systems shrink in size. Restricted geometries in the mesoscopic and nanoscales have become ubiquitous eg., nanochannels, quantum wires, thin films and quantum dots, to name a few. Quantum mechanics kicks in at sufficiently small confining lengths via quantization of various physical quantities characterizing  the system, referred to as the quantum size effect(QSE). The influence on the system of surface roughness of confining walls  is a more complicated, but fundamentally  important problem.  I address this problem and analyse transport properties of a Fermi liquid mesoscopically confined by rough walls. 

Liquid $^3$He is a paradigm for Fermi liquids and exhibits a rich superfluid phase diagram with a complex order parameter having unconventional  $p$-wave pairing symmetry \cite{VollhardtWolfle}.  
In the absence of an applied magnetic field, there are two stable bulk superfluid phases with triplet pairing. The B-phase is a timereversal-invariant phase with an isotropic gap and is the stable low-temperature phase. The A-phase is a chiral phase with nodes in the gap and an intrinsic nonzero pair angular momentum. Superfluid $^3$He is a topological superfluid and the existence of edge currents and surface excitations have been predicted in both A and B phases\cite{VolovikBook}. These Majorana states have unique non-local properties that render them suitable to applications in quantum computing\cite{MajoranaQuantumComputationPRL}. The detection of these states has however been elusive\cite{ChungZhangPRL2009}. Superfluid $^3$He provides a model system in which to search for these states.  With the development in the fabrication of nanocavities suitable to study superfluid $^3$He, the search for the Majorana states in this system has intensified in confined geometries more recently \cite{SaundersPRL2013,LevitinScience2013}.
The study of confined liquid $^3$He has also been fundamentally driven by the prediction of 
 new phases not present in the bulk  \cite{VorontsovSauls}. Confinement and size effects inevitably raise the questions of scattering conditions at the surfaces and the effects of surface roughness.
 
Earlier studies of the effect of surface roughness in liquid $^3$He slabs have reported anomalous relaxation rates of the Fermi liquid in a torsion oscillator\cite{Casey2004}. In this paper, I present a rigorous calculation of heat and momentum transport of normal liquid $^3$He in a slab geometry and obtain expressions for the viscosity and thermal conductivity in this anomalous regime, as a function of the surface roughness. These calculations propose a method to characterise  the normal state of liquid $^3$He in a slab; a crucial step in unravelling the mysteries of the confined superfluid phase diagram. These results are analogously applicable to metallic thin film systems, where the electronic Fermi liquid can be expected to show anomalous behaviour in transport properties along the same lines as reported in this paper.

Transport in the bulk quantum fluid, $^3$He is well-understood within the framework of Landau's Fermi liquid theory \cite{LandauLifshitz} for temperatures less than a few hundred milliKelvin. The relevant excitations of the system that determine the physics in the Fermi-liquid regime are quasiparticles with well-defined momentum, $\vec{\bf{p}}_F$ and energy, $\varepsilon_F$ (Fermi momentum and energy, respectively). At temperatures below $100$mK (for $T\ll \varepsilon_F/k_B \sim 1K$ for liquid $^3$He), the quasiparticles form a system of degenerate fermions with effective mass $m$ i.e., $\varepsilon_F = p_F^2/2m$ with Fermi-liquid interactions described in terms of well-known Landau parameters\cite{LeggettRevModPhys,WheatleyRevModPhys}. Quasiparticles scatter with each other through inelastic binary quasiparticle collisions  with a scattering rate $\propto T^2$. Transport of heat and momentum is mediated by this quasiparticle scattering mechanism and the viscosity $\eta_b\propto T^{-2}$ and the thermal conductivity $\kappa_b \propto T^{-1}$ in bulk liquid $^3$He. The inelastic mean free path, $\ell_{in}\propto T^{-2}$ and ranges from a few nanometers at high temperatures up to tens of microns at low temperatures. As the size of the system reduces and becomes comparable to $\ell_{in}$, transport properties enter the Knudsen regime. For liquid $^3$He, since $\ell_{in}$ spans three orders of magnitude on cooling, a wide range of system sizes can be sampled in the Knudsen limit. For smaller systems, the QSE operates in this limit and the roughness of the confining surfaces begins to play a role in the physics. Inelastic quasiparticle scattering events are fewer and are mediated by elastic scattering off the rough surface. On approaching the Knudsen limit, quasiparticles maintain coherence while undergoing both scattering processes and transport is determined by quasiparticles experiencing a coherent "mixed" scattering channel. This is the regime of interest in the rest of this paper.

When Fermi liquid $^3$He is confined to a thin slab of thickness $L$, size quantization in the transverse direction splits the quasiparticle spectrum into a set of bands $\varepsilon_n(\vec{\bf{q}})$, with quasiparticles in each band moving freely with longitudinal (in-plane) momentum $\vec{q}$. For $L\lesssim \ell_{in}$, this modifies the phase space available for inelastically scattered quasiparticles and hence the binary quasiparticle scattering rate. For a rough substrate surface with small roughness given by a height profile, $h(x,y)$, ($z$ being the confinement direction), quasiparticles experience a local confinement spectrum that varies in the longitudinal direction. For $h/L\ll 1$, they may be viewed as free quasiparticles making transitions between quantized bands as they move along their trajectory, $\hat{\bf{p}}_F$. This picture can be formalized in terms of a virtual disorder potential in the bulk, that drives these transitions in a system with flat walls. Such a transformation was suggested independently by Tesanovic {\it{et. al}} \cite{Tesanovic} and by Trivedi and Ashcroft \cite{TrivediAshcroft} and employed to formulate a theory for a layered system by Meyerovich and coworkers \cite{Meyerovich98,Meyerovich99,Meyerovich2002}. 
The scattering off the virtual disorder potential is found to be $\propto (h/L)^2 \ll k_B T\sim$ the excitation energy of a quasiparticle. Hence, it may be assumed that the elastic scattering off the roughness does not affect the intermediate states that mediate the inelastic quasiparticle collisions. The phase space available for scattered quasiparticles undergoing binary inelastic collisions, is given by the quasiparticle density of states, which can be expanded in $h/L$ for $h/L\ll 1$, and for a randomly rough surface with $\langle h(x,y) \rangle_{(x,y)} = 0$ (angular brackets denotes surface average). The scattering rate for inelastic collisions in this system is found to be $\propto T$, and can be expressed\cite{Meyerovich2000} in terms of the surface roughness power spectrum $\langle h(\vec{\bf{q}})h(\vec{\bf{q}}\,')\rangle_{(x,y)} \equiv \zeta(\vec{\bf{q}}-\vec{\bf{q}}\,')$, often referred to as the autocorrelation function of the surface roughness. This anomalous linear temperature dependence of the relaxation rate has been measured \cite{Casey2004} in a torsional oscillator for $^3$He films of thickness a  few  hundred $nm$s  and interpreted successfully by the Meyerovich formulation \cite{RoughnessPRL}. Consequently, transport properties are expected to exhibit this anomalous temperature dependence and have been investigated theoretically in this paper. 

The calculation presented here predicts an anomalous temperature dependence of the thermal conductivity and viscosity of $^3$He in thin films with thickness in the range of $100$nm to $\sim 1\mu$m. The transport coefficients are given by expressions derived using the formulation discussed above, and can be calculated numerically for given surface roughness autocorrelation function. The latter can be independently determined experimentally using scanning probe techniques, and goes into the calculation as an {\it{ab initio}} input. The calculation of transport coefficients in these terms is especially effective in resolving the effects of confinement versus the effects of surface roughness, and has not been done before.

\section{Transport Theory}
\noindent
Consider Fermi liquid $^3$He confined to an infinite slab of thickness $L$, with one rough wall having surface roughness power spectrum $\zeta(\vec{\bf{q}}-\vec{\bf{q}}\, ')$, the second wall being smooth. For isotropic randomly rough surfaces, $\zeta = \zeta(\mid\vec{\bf{q}}-\vec{\bf{q}}\, '\mid)$. This is the autocorrelation of the surface roughness viz., if the height profile of the surface roughness is $h(x,y)\equiv h(\vec{\bf{s}})$, then
\begin{equation}
\label{defn-zeta}
\zeta(\vec{\bf{q}})=\int\nolimits d\vec{\bf{s}}\, e^{i\vec{\bf{q}}\cdot\vec{\bf{s}}/\hbar}\,\zeta(\vec{\bf{s}})\,\,\,;\,\,\, 
\zeta(\vec{\bf{s}}) = \int\nolimits d\vec{\bf{s}_1} h(\vec{\bf{s}_1})\,h(\vec{\bf{s}}+\vec{\bf{s}_1})\,\,\,,
\end{equation}
$\vec{\bf{q}}$ and $\vec{\bf{s}}$ are two-dimensional in-plane vectors in the momentum and real spaces, respectively ($xy$ plane as shown in Fig.$1$). As discussed in the preceding paragraphs, scattering of $^3$He quasiparticles off the surface roughness can be formulated in terms of scattering off a virtual disorder potential in the bulk in a geometric system with flat walls \cite{Tesanovic}. This is achieved by a coordinate transformation, first suggested by Trivedi and Ashcroft \cite{TrivediAshcroft} which applies for inhomogeneities large on the scale of the quasiparticle wavelength, $k_F h\gg 1$ and small on the scale of system size, $h\ll L$. For weak roughness, $h\ll\ell_{in}$ in the Knudsen regime and the system can be treated in the continuum limit  $k_F L\gg1$. Of course, the mapping transformation applies when the stretching of coordinates as effected by the transformation is such that $h$ is smaller than the distance over which $^3$He quasiparticles maintain coherence, or the length over which quasiparticle decohere, given by the thermal rate $\sim \hbar/k_BT$.i.e., $h\ll v_F \hbar/k_B T\sim 50$ nm at $P=0$bar and $T=10$mK.

The relaxation rate is given by \cite{Meyerovich2000}
\begin{equation}
\label{equation1}
\frac{1}{\tau_{eff}^j(\vec{\bf{p}})} = \frac{1}{\tau_b^j(\vec{\bf{q}})} + \Sigma_{j'=1}^{S}\int\nolimits\frac{W_{jj'}(\vec{\bf{q}},\vec{\bf{q}}\,')/\tau_b^{j'}(\vec{\bf{q}}\,')}{(\varepsilon_{j'}(\vec{\bf{q}}\,')-\varepsilon_F)^2/\hbar^2 + (1/2\tau_b^{j'}(\vec{\bf{q}}\,'))^2}\,\,\frac{d\vec{\bf{q}}\,'}{(2\pi\hbar)^2}\,\,\,\,\,\,\,,
\end{equation}
to order  $\vartheta((h/L)^2)$,  the leading term in a $h/L\ll 1$ expansion. Here $\tau_b$ is the bulk relaxation rate including inelastic scattering processes, $j, j'$ are the band indices for the quantized minibands in the $z$-direction of confinement, and $\varepsilon_F$ is the Fermi energy. $S$ is the total number of minibands that is chosen to be summed over. The scattering probability is given by
\begin{equation}
\label{W}
W_{jj'}(\vec{\bf{q}},\vec{\bf{q}}\,') = \frac{\pi^4\hbar^2}{m^2 L^6}\,\,\zeta(\vec{\bf{q}}-\vec{\bf{q}}\, ')\,\,j^2j'^2\,\,\,\,\,\,\,.
\end{equation}
Here, $m$ is the quasiparticle mass. For a slab with both surfaces being randomly rough, another term of similar form with the surface roughness power spectrum $\zeta_2$ of the second surface adds to the probability in the expression above. If the surface roughnesses of both surfaces are in turn correlated, then the simple formulation above no longer holds. This regime is much more complicated and has been addressed by Meyerovich \cite{Meyerovich98}. However, all following arguments assume uncorrelated surface roughness away from this "quantum resonance" regime.


In the normal state, on application of a thermal gradient $\nabla{T}$ or fluid flow $\vec{\bf{u}}$, the Fermi liquid density of states remains unchanged and the quasiparticle distribution function responds to the applied gradient/flow by a change $\delta n_{\vec{\bf{p}}\sigma}$. The linearized Boltzmann-Landau transport equation for the distribution function $n_{\vec{\bf{p}}\sigma}$ may be used for quasiparticles of momentum $\vec{\bf{p}}$ and spin $\sigma$, in the steady state viz.,
\begin{equation}
\label{Boltzmann_equation}
\frac{\partial n_{\vec{\bf{p}}\sigma}}{\partial\varepsilon_{\vec{\bf{p}}\sigma}}\,\vec{\bf{v}}_{\vec{\bf{p}}\sigma}\cdot\nabla(\delta\varepsilon_{\vec{\bf{p}}\sigma}) = I[n_{\vec{\bf{p}}\sigma}]\,\,\,\,\,.
\end{equation}
The driving term, $I[n_{\vec{\bf{p}}\sigma}]$ is the collision integral given by inelastic scattering as well as scattering from the virtual disorder potential at a rate set by equation(\ref{W}). Consider linear deviations from equilibrium viz.,
\begin{equation}
n_{\vec{\bf{p}}} = n_{\vec{\bf{p}}}^0 +\delta n_{\vec{\bf{p}}}\,\,\,.
\end{equation}
Then
\begin{equation}
\label{delta_n1}
n_{\vec{\bf{p}}}(1-n_{\vec{\bf{p}}\,'}) - n_{\vec{\bf{p}}\,'}(1-n_{\vec{\bf{p}}}) = \delta n_{\vec{\bf{p}}} - \delta n_{\vec{\bf{p}}\,'}
\end{equation}
with
\begin{equation}
\label{n0_def}
n_i^0 = \frac{1}{1 + e^{\beta\varepsilon_i}}\,\,\,\,\,; \,\,\,\,\,\beta \equiv (k_BT)^{-1}\,,
\end{equation}
the Fermi distribution function. Therefore,
\begin{equation}
\label{n0}
\frac{\partial n_i^0}{\partial\varepsilon_i} = -\frac{\beta e^{\beta\varepsilon_i}}{(1 + e^{\beta\varepsilon_i})^2} = -\beta n_i^0 (1 - n_i^0)\, = \frac{-\beta}{4} sech^2(\chi/2).
\end{equation}
Here $\chi = \frac{\varepsilon-\varepsilon_F}{k_BT}$. Define a function $\Phi_i$ such that
\begin{equation}
\label{delta_n2}
\delta n_i = - \frac{\partial n_i^0}{\partial \varepsilon_i}\Phi_i\,\,\,\,\,\,\,\left(i=\vec{\bf{p}},\vec{\bf{p}}'\right)\,\,\,.
\end{equation}
The left hand side of the Boltzmann equation (\ref{Boltzmann_equation}) can now be worked out using equations(7-9), and with the collision integral derived in the following section, the Boltzmann equation reduces to an equation for $\Phi_i$.

\subsection{Collision integral}

The right hand side of the Boltzmann equation is the driving term given by the collision integral. Consider the roughness-induced term viz., the second term in equation(\ref{equation1}) in order to formulate the collision integral. The first term in equation(\ref{equation1}) is the bulk term and is added onto the wall-driven term using a relaxation time approximation viz., $\eta_{eff}^{-1} = \eta_{bulk}^{-1} + \eta_{wall}^{-1}$, and similarly for the thermal conductivity. Linearizing the collision integral, using equation (\ref{W}) and going to the continuum limit (as discussed in \cite{Meyerovich2000}) viz., $p_zL\gg 1$ and $j\gg 1$, 
\begin{equation}
\label{coll_int1}
I[n_{\vec{\bf{p}}\sigma}] 
= -2L\int\nolimits \frac{d\vec{\bf{p}}\,'}{(2\pi\hbar)^3} \,\frac{\zeta(\vec{\bf{q}}-\vec{\bf{q}}\,')}{m^2L^2\hbar^2\tau_b}\,\frac{p_z^2p_z'^2}{(\chi\,'^2 k_B^2 T^2/\hbar^2 + (2\tau_b)^{-2})}\,\delta (\chi-\chi')(\delta n_{\vec{\bf{p}}}-\delta n_{\vec{\bf{p}}\,'})
\end{equation}
with $\chi = \frac{\varepsilon-\varepsilon_F}{k_BT}$ and $p_z = \frac{j\pi\hbar}{L}$. Using equations(\ref{n0} and \ref{delta_n2}), the right hand side of the Boltzmann equation above can be worked out in terms of $\Phi_i$. The deviation $\delta n_{\vec{\bf{p}}\sigma}$ can then be worked out by solving the Boltzmann-Landau equation for $\Phi_{\vec{\bf{p}}\sigma}$.

\subsection{Viscosity}
Let the fluid (flow) velocity, $\vec{\bf{u}}$ be along the $x$-direction.i.e., $\vec{\bf{u}}=u(z)\hat{\bf{x}}$. The geometry is illustrated in Fig.$1$. Clearly, $x$ and $y$ are equivalent directions. The momentum flux tensor is given by the dissipative part of the stress tensor $\sigma_{xz}$
\begin{equation}
\label{viscosity-defn}
\sigma_{xz} \equiv \eta\frac{\partial u_x}{\partial z} = -\Sigma_{\sigma}\int\nolimits \frac{d^3p}{(2\pi\hbar)^3}\,p_{x\sigma}\,(\vec{\bf{v}}_{\vec{\bf{p}}\sigma})_z\,\delta n_{\vec{\bf{p}}\sigma}\,\,\,,
\end{equation}
where $\eta$ is the flow viscosity.
With the fluid flow,
\begin{eqnarray}
\label{gradE}
\varepsilon_{\vec{\bf{p}}} &=& \varepsilon_F + \vec{\bf{p}}\cdot\vec{\bf{u}}
\nonumber
\\
\nabla\varepsilon_{\vec{\bf{p}}} &=& \nabla(p_x u_x) = \hat{\bf{z}} p_x \frac{\partial u}{\partial z}\,\,\,.
\end{eqnarray}
Plugging this into the Boltzmann equation(\ref{Boltzmann_equation}), and using equation(\ref{n0}),
\begin{equation}
\label{Boltzmann-equation2}
\beta\,n_{\vec{\bf{p}}}^0 (1-n_{\vec{\bf{p}}}^0)\,(\vec{\bf{v}}_{\vec{\bf{p}}})_z\,p_x\,\frac{\partial u_x}{\partial z} = I[n_{\vec{\bf{p}}\,'}]
\end{equation}
for spin independent scattering.
Putting equations(\ref{n0} and \ref{delta_n2}) in the collision integral (equation (\ref{coll_int1})),
\begin{equation}
\label{coll_int2}
I[n_{\vec{\bf{p}}}] = \frac{-2 N_f p_z^2}{\hbar^2 m^2 L \tau_b} \frac{n_{\vec{\bf{p}}}^0(1-n_{\vec{\bf{p}}}^0)}{\left((\varepsilon-\varepsilon_F)^2/\hbar^2 + (1/2\tau_b)^2\right)}\,\int\nolimits\frac{d\Omega'}{4\pi}\zeta(\vec{\bf{q}}-\vec{\bf{q}}\,')\,p_z'^2\,(\Phi_{\vec{\bf{p}}} - \Phi_{\vec{\bf{p}}\,'})
\end{equation}
where $N_f$ is the density of states at the Fermi level, $N_f = m p_F/2\pi^2\hbar^3$.   Assume  the momentum and energy dependences are separable and let
\begin{equation}
\label{def_Phi}
\Phi_{\vec{\bf{p}}} \equiv \varphi(\vec{\bf{p}})\,\psi(\chi)\,\,\,\,\,.
\end{equation}
This is a standard assumption in the linear regime of the Boltzmann equation and is justified considering the evolution of the distribution function and its deviations from its equilibrium momentum-independent form.
Use the ansatz
\begin{equation}
\label{psi_ansatz}
\psi(\chi) = \frac{-\hbar^2 L m^2 \tau_b}{2 N_f p_F^4 (k_B T)}\,\left(\chi^2\frac{k_B^2 T^2}{\hbar^2} + (\frac{1}{\tau_b})^2\right)\,\,\,
\end{equation}
and the Boltzmann equation reduces to an integral equation for $\varphi(\vec{\bf{p}})$,
\begin{equation}
\label{phi-eqn1}
\frac{p_F^2}{m}sin\theta\,cos\theta\,cos\phi\frac{\partial u}{\partial z} = cos^2\theta\int\nolimits\frac{d\Omega'}{4\pi}\zeta(\vec{\bf{q}}-\vec{\bf{q}}\,')cos^2\theta'(\varphi(\theta,\phi)-\varphi(\theta',\phi'))\,
\end{equation}
with $\vec{\bf{p}}=(\vec{\bf{q}},p_z) = (psin\theta cos\phi, psin\theta sin\phi, p cos\theta)$ and analogously for $\vec{\bf{p}}\,'(\theta\,',\phi\,')$. Here, $d\Omega' = d (cos\theta') d\phi'$ and the integral is over the direction $\hat{\bf{p}}\,'$. 
\noindent
Define
\begin{equation}
\bar{\varphi}(\theta,\phi) = \varphi(\theta,\phi)\frac{1}{(p_F^2/m)\cdot\partial u/\partial z}\,\,\,.
\end{equation}
The solution for the viscosity is then given by the following equation, accompanied with a self-consistency equation for $\bar{\varphi}$ (from equation(\ref{phi-eqn1})).
\begin{equation}
\label{eta2}
\eta = \frac{-\hbar^2 L \tau_b}{2k_BT}\left(\pi^2\frac{k_B^2T^2}{3 \hbar^2} + (\frac{1}{2\tau_b})^2\right)\int\nolimits \frac{d\Omega}{4\pi} \,sin\theta\,cos\theta\,cos\phi\,\bar{\varphi}(\theta,\phi)
\end{equation}
\begin{equation}
\label{eta2-phi}
\bar{\varphi}(\theta,\phi)\,cos^2\theta\int\nolimits\frac{d\Omega'}{4\pi}\zeta(\vec{\bf{q}}-\vec{\bf{q}}\,')\,cos^2\theta' = sin\theta\,cos\theta\,cos\phi + cos^2\theta\int\nolimits\frac{d\Omega'}{4\pi}\zeta(\vec{\bf{q}}-\vec{\bf{q}}\,')\,cos^2\theta'\,\bar{\varphi}(\theta',\phi')
\end{equation}
Expand $\bar{\varphi}$ in the spherical harmonics $Y_{lm}$ :
\begin{equation}
\label{spherical_harmonics_expansion}
\bar{\varphi}(\theta,\phi) = \Sigma_{lm}\,Y_{lm}(\theta,\phi) \bar{\varphi}_{lm}
\end{equation} 
and
\begin{equation}
\label{Ieta}
I_{\eta}\equiv\int\nolimits\frac{d\Omega}{4\pi}  \,sin\theta\,cos\theta\,cos\phi\,\bar{\varphi}(\theta,\phi) = \frac{1}{\sqrt{15}}(\bar{\varphi}_{2,-1} - \bar{\varphi}_{2,1})
\end{equation}
using the orthogonality of the spherical harmonics.
Putting this back in equation(\ref{eta2}), 
\begin{equation}
\label{eta3}
\eta = \frac{\hbar^2 L \tau_b}{2k_BT}\left(\pi^2\frac{k_B^2T^2}{3 \hbar^2} + (\frac{1}{2\tau_b})^2\right)\left(\bar{\varphi}_{2,1} - \bar{\varphi}_{2,-1}\right)\frac{1}{\sqrt{15}}
\end{equation}
is the expression to be evaluated for $\eta$ along with the self-consistency equation(\ref{eta2-phi}) for $\bar{\varphi}$. 

Examining the symmetry (with respect to $lm$) of the Boltzmann equation, the only non-zero components of both sides of the equation for $\bar{\varphi}$, viz., equation(\ref{phi-eqn1}), are the $l=2\,;m=\pm 1$ components. Examining the angular moments in the integral on both sides of equation(\ref{phi-eqn1}), the formulae in Appendix D can be used to deduce $\bar{\varphi}_{lm} = 0\,;m\neq \pm1$ and $\bar{\varphi}_{l\pm 1} = 0 \,\forall\,l\neq2,4,6$. Therefore,  a complete expansion of $\bar{\varphi}$ to all orders is given by
\begin{eqnarray}
\bar{\varphi}(\theta,\phi) = \bar{\varphi}_{21}\,Y_{2,1}(\theta,\phi) &+&  \bar{\varphi}_{2-1}\,Y_{2,-1}(\theta,\phi) +  \bar{\varphi}_{41}\,Y_{4,1}(\theta,\phi) 
\nonumber
\\
&+&  \bar{\varphi}_{4-1}\,Y_{4,-1}(\theta,\phi) +  \bar{\varphi}_{61}\,Y_{6,1}(\theta,\phi) +  \bar{\varphi}_{6-1}\,Y_{6,-1}(\theta,\phi)\,\,\,.
\end{eqnarray}

If this expression is truncated after the first two terms,
the viscosity is  given by the following expression,
\begin{equation}
\label{eta4}
\eta = \frac{\sqrt{2}}{15}\,L\,\tau_0(P)\,k_B\pi^2 \mathcal{G}_{\eta}\left(\frac{1}{T} + \frac{3 \hbar^2}{4\pi^2k_B^2}\frac{T}{4\tau_0^2}\right)\varpi\,\,\,,
\end{equation}
with $\tau_b = \tau_0(P)/T^2$, $P$ being pressure, $\mathcal{G}_{\eta} = \frac{1}{3}$ is a numerical constant, and  $\varpi$ is a geometric factor given by various components of the roughness structure factor, $\zeta$ (Appendix B).
By symmetry (and trivially by explicit calculation), $\sigma_{xz} = \sigma_{yz} \equiv \sigma_{||}$ and the result for $\eta$ in both parallel directions are identical. 

{\subsection{Thermal Conductivity}

The thermal conductivity is calculated from the Boltzmann equation(\ref{Boltzmann_equation}), with the collision integral given by the scattering off the surface roughness, equation(\ref{coll_int1}). As in the case of viscosity, use the function $\Phi$ as defined in equation(\ref{delta_n2}), with  $\Phi(\vec{\bf{p}}) = \varphi(\vec{\bf{p}})\,\psi(\chi)$. Of course, the functions $\varphi(\vec{\bf{p}})$ and $\psi(\chi)$ are different from their values calculated in the viscosity case. 
For this case, try the ansatz
\begin{equation}
\psi(\chi) = \frac{-\hbar^2\,Lm^2\tau_b}{2N_f\,p_F^4\, k_B T}\left(\frac{\chi^2k_B^2T^2}{\hbar^2}+(\frac{1}{2\tau_b})^2\right)\chi\,\,\,.
\end{equation}

Consider a temperature gradient in the parallel direction, $\nabla T\|\hat{\bf{x}}$}.
This is the case of natural physical interest with the temperature gradient being in the plane of the slab, viz., the $xy$ plane. The thermal conductivity $\Bar{\kappa}$ is the linear response to a thermal gradient defined thus
\begin{equation}
\Bar{\kappa}\cdot\nabla T = \Sigma_{\sigma}\int\nolimits\frac{d^3p}{(2\pi\hbar)^3}(\varepsilon_{\vec{\bf{p}}\sigma}-\mu)\,(\vec{\bf{v}}_{\vec{\bf{p}}\sigma})\delta n_{\vec{\bf{p}}\sigma}\,\,\,.
\end{equation}
In this case,
\begin{equation}
\kappa_{xx}|\nabla T| = 2\int\nolimits\frac{d^3p}{(2\pi\hbar)^3}\,(\varepsilon_{\vec{\bf{p}}}-\mu)\,(\vec{\bf{v}}_{\vec{\bf{p}}})_x\delta n_{\vec{\bf{p}}}\,\,\,.
\end{equation}
Plugging this into the Boltzmann equation, obtain an integral equation for $\bar{\varphi}$,
\begin{equation}
\label{phi_kappa}
\bar{\varphi}(\vec{\bf{p}})\,cos^2\theta\int\nolimits\frac{d\Omega'}{4\pi}\,\zeta(\vec{\bf{q}}-\vec{\bf{q}}\,')\,cos^2\theta' = sin\theta\,cos\phi + cos^2\theta\int\nolimits\frac{d\Omega'}{4\pi}\zeta(\vec{\bf{q}}-\vec{\bf{q}}\,')\,cos^2\theta'\,\bar{\varphi}(\vec{\bf{p}}\,')\,\,\,,
\end{equation}
with $\bar{\varphi}(\vec{\bf{p}}) = \frac{\varphi(\vec{\bf{p}})}{k_B v_F|\nabla T|}$. $\kappa_{xx}$ is given by 
\begin{equation}
\label{kxx}
\kappa_{xx} = \frac{L\tau_bk_B\pi^2}{6 k_F^2}\left(\frac{8}{5}\pi^2\frac{(k_BT)^2}{\hbar^2} + (\frac{1}{2\tau_b})^2\right)\int\nolimits\frac{d\Omega}{4\pi}\,sin\theta\,cos\phi\,\bar{\varphi}(\vec{\bf{p}})\,\,\,.
\end{equation}
Expanding $\bar{\varphi}$ in spherical harmonics as in equation(\ref{spherical_harmonics_expansion}) before, evaluate
\begin{equation}
I_{\kappa}^{xx}\equiv\int\nolimits\frac{d\Omega}{4\pi}\,sin\theta\,cos\phi\,\bar{\varphi}(\vec{\bf{p}}) = \frac{1}{\sqrt{6}}\,(\bar{\varphi}_{1,-1} -\bar{\varphi}_{1,1})\,\,\,.
\end{equation}
and get an expression for the thermal conductivity  $\kappa_{xx}$,
\begin{equation}
\label{kxx2}
\kappa_{xx} = \frac{L\,\tau_bk_B\pi^2}{6\sqrt{6}\,k_F^2}\left(\frac{8}{5}\pi^2(\frac{k_BT}{\hbar})^2 + (\frac{1}{2\tau_b})^2\right)(\bar{\varphi}_{1,-1}-\bar{\varphi}_{1,1})\,\,\,.
\end{equation}
Equations(\ref{phi_kappa} and \ref{kxx2}) need to be solved simultaneously to get $\kappa_{xx}$.
The driving term in the equation for $\bar{\varphi}$ is the first term on the right hand side of equation(\ref{phi_kappa}). For the $\kappa$-case, this term has $l=1,m=\pm 1$ symmetry and hence, only $lm=1\pm 1$ components of both right and left hand side terms will be nonzero. Examining the  moments in the integral on both sides, $\bar{\varphi}_{lm} = 0 ; m\neq 1$ and $\bar{\varphi}_{l\pm 1} = 0 \,\forall\, l\neq 1,3,5$. Therefore, a full and complete expansion of $\bar{\varphi}$ to all orders is
\begin{eqnarray}
\bar{\varphi}(\theta,\phi) = \bar{\varphi}_{11}\,Y_{1,1}(\theta,\phi) &+&  \bar{\varphi}_{1-1}\,Y_{1,-1}(\theta,\phi) +  \bar{\varphi}_{31}\,Y_{3,1}(\theta,\phi) 
\nonumber
\\
&+&  \bar{\varphi}_{3-1}\,Y_{3,-1}(\theta,\phi) +  \bar{\varphi}_{51}\,Y_{5,1}(\theta,\phi) +  \bar{\varphi}_{5-1}\,Y_{5,-1}(\theta,\phi)\,\,\,.
\end{eqnarray}
If the expression above is truncated after the first two terms,
the thermal conductivity is  given by the following expression
\begin{equation}
\label{kxx3}
\kappa_{xx} =  \frac{L k_B^3 \pi^4}{\hbar^2k_F^2}\,\tau_0(P)\,\mathcal{G}_{\kappa}(1 + \frac{5T^2}{64\pi^2\tau_0^2}\,\frac{\hbar^2}{k_B^2})\varpi_{\kappa}\,\,\,,
\end{equation}
with $\tau_b = \tau_0(P)/T^2$ as before, $\mathcal{G}_{\kappa} = \frac{4}{45}$ is a numerical constant and  $\varpi_{\kappa}$ is a geometric factor given by various components of the roughness structure factor, $\zeta$ (Appendix C).
The microscopic processes responsible for heat transfer are the processes of quasiparticle scattering and scattering off the virtual disorder potential. Since these processes are isotropic at each scattering event, the transverse components of heat transfer cancel and $\kappa_{xz}=\kappa_{xy}=0$. If $\nabla T\|\hat{\bf{y}}$, by symmetry (and trivially by explicit calculation), $\kappa_{xx}=\kappa_{yy} \equiv \kappa_{||}$ and $\kappa_{yz} = \kappa_{xz} = \kappa_{yx} =0$. 

\subsection{Low Temperature Limit}
\noindent
As $T\rightarrow 0$, the inelastic scattering freezes out and  consequently, the scattering process considered in the calculation discussed above freezes out. In this limit, scattering is purely from  the roughness and this is a strictly elastic channel that has not be included in the considerations thus far. It is temperature-independent and the elastic scattering rate, $\tau_{el}^{-1}\sim v_F/\ell_{surface}$ where $\ell_{surface}$ is a characteristic length set by the surface roughness, estimated by Tesanovic {\it{et. al.}} {\cite{Tesanovic}}. $\ell_{surface}$ arises from residual roughness scattering and for a white noise autocorrelation function with rms roughness $\Delta h\sim 10$nm, $\ell_{surface}\sim10^{-5}m$. Clearly, $\ell_{surface}\gg \ell_{in}$, the inelastic mean free path in the bulk. As $T\rightarrow 0$, $\eta_{T\rightarrow 0}\sim 3 X 10^{-4}$kg m$^{-1}$ sec$^{-1}$ at $P=0$bar and saturates to a value of this order at low temperatures. This is the residual viscosity. The residual $\kappa_{T\rightarrow 0} \propto T$.  These residual values depend upon pressure and roughness as do the cross-over temperatures to the residual behaviour in the low temperature limit.

The residual value of the transport comes from the quantum size effect and has no classical analogue. In the classical limit, the transport vanishes in the low temperature limit as a beam of ballistic quasiparticles  can propagate parallelly through the film with zero effect from the roughness in the absence of any bulk relaxation mechanism. However, in the quantum-mechanical case, the quantum mechanical zero-point motion excludes strict two dimensional confinement in the film plane and gives rise to nonzero transport in the residual limit. In general $\ell_{surface}$ is the length scale that corresponds to the relaxation rate from surface scattering only i.e., in the absence of all other scattering channels at $T=0$. It is the mean free path of the residual scattering rate arising from quantum size effects and surface roughness scattering. It sets a length scale over which the quasiparticle wave function decoheres in this limit.

\subsection{Crossovers}

If the inelastic quasiparticle scattering rate is very large, then inelastic processes dominate and wash out the anomalous effect that arises when inelastic events are mediated by elastic scattering off the virtual disorder potential.  Bulk behaviour dominates at temperatures  $T>T^{\star}= k_B \tau_0(P)/\hbar \sim 200mK$ at $P=0$bar. 

If the inelastic scattering rate is very small and $\ell_{in}\gg\ell_{el}$, the characteristic length scale for elastic scattering (mediated by inelastic events) off the virtual disorder potential, then inelastic quasiparticle scattering events are no longer mediated by elastic events and the scattering will be dominated by the scattering from surface roughness. The anomalous effect disappears then for $T<T^{\star}\,'\sim\tau_0(P)\,k_F(h/L)\sim 1mK$ at $P=0$ bar.

Therefore, the anomalous effect only exists in a regime $T^{\star}\,'<T<T^{\star}$ , where the limiting values depend on pressure, slab thickness and roughness.

\section{Surface Roughness}

The roughness of confining walls can be determined by high resolution surface microscopy techniques. The structure factor of roughness is treated as known for  purposes of this work and goes into the calculation as a fixed input derived from experiment. For purposes of discussing results of the theoretical predictions, we use three varied autocorrelation functions. The first is a Gaussian autocorrelation function with $\zeta(\vec{\bf{q}})_G = 2\pi l^2R^2\,e^{-|\vec{\bf{q}}|^2R^2}$, where $l$ and $R$ refer to Gaussian parameters reflecting height and correlation length of inhomogeneities. 
The second line shape is that of a fractal autocorrelation function, $\zeta_F(\vec{\bf{q}}) = (\Delta h)^2/|\vec{\bf{q}}|^H$ of a self-affine fractal, $\Delta h$ being the rms roughness and $H$ the fractal exponent.
The third type of roughness used is the white noise spectrum $\zeta_{WN} = (\Delta h)^2/k_F^2$, with $\Delta h$ being the rms roughness. 
These autocorrelation function types have been chosen to match with those that are found on real rough surfaces \cite{Persson}. The values of the parameters chosen illustrate the typical values that correspond to smooth polished surfaces as may be used in a typical $^3$He slab experiment.

\section{Results and Discussion}

The calculated values of viscosity in the temperature range where anomalous behaviour is expected  are shown in Fig.$2$. For weaker roughness ($\zeta_{WN}$ and $\zeta_F$ in Fig.$2$), the crossover from bulk $\eta_b \propto 1/T^2$ to the linear regime $\eta_{||}\propto 1/T$ is noticeable in the temperature range of validity of the formulation discussed. For rougher surfaces, (eg., $\zeta_G$ in Fig.$2$), the wall-component dominates and the $\eta_{||}\propto 1/T$ behaviour spans a large temperature range. The shape of the autocorrelation function plays a significant part in determining the strength of roughness in a sense, as can be seen by comparing $\zeta_{WN}$ and $\zeta_G$ with the same rms roughness in Fig.$2$. The shape of $\zeta$ enters the expression for $\eta$ implicitly via the $\bar{\varphi}$ function, in equation(\ref{eta3}), and reflects the dependence of the spatial form of roughness correlation (or lack of it) on momentum transfer. Wall scattering events that mediate inelastic quasiparticle collisions enhance momentum transfer in the parallel direction and lead to a reduction in viscosity with respect to the bulk.

The calculated values for heat transport in a similar temperature range are also shown in Fig.$2$. Wall scattering events that mediate quasiparticle collisions supress heat transfer and on cooling, the thermal conductivity saturates to a less effective value compared to the bulk.

\section{Summary}

I have presented theoretical calculations of the viscosity and thermal conductivity in a regime where transport is dominated by a coherent "mixed" scattering channel of inelastic quasiparticle collisions mediated by elastic scattering events off the surface roughness of confining walls. Thermal conduction which is highly effective in bulk $^3$He is suppressed and saturates to a residual value in this regime. Momentum transfer is rendered more effective in this regime and the viscosity is suppressed with respect to the bulk. The crossover to this regime on cooling from the bulk is a function of pressure, film thickness and surface roughness, with both the size and form of the autocorrelation function of roughness being considerable in this context. This anomalous transport should be observable in films of thickness of a few hundred $nm$s in the temperature range $\sim 10-200$mK, the effects being more pronounced at higher temperatures in thinner films. The effect of  residual scattering at low temperatures might affect the phase diagram in the superfluid state.

\section*{Acknowledgements}

I would like to thank the Centre for NanoScience and Engineering, IISc for its support.

\appendix
\begin{center}
\section{AutoCorrelation Function}
\end{center}
The autocorrelation function, $\zeta$ is a two-dimensional quantity that depends on two-dimensional in-plane vectors, $\vec{\bf{q}}$ and $\vec{\bf{q}}\,'$. With $\phi$ being the azimuthal angle, $\zeta$ may be expanded as $\Sigma_m\zeta_m(\theta,\theta')\,e^{im\phi}\,e^{-im\phi'}$. With the in-plane isotropy of $\zeta$ in two dimensions, it is simply an expansion in $\phi-\phi'$. The coefficients of this expansion depend upon the magnitudes of $\vec{\bf{q}}$ and $\vec{\bf{q}}\,'$. viz., on $\theta$ and $\theta'$.  By the trivial symmetry of the function perpendicular to the surface (since $\zeta$ is specified as the autocorrelation function only on the surface), the aformentioned coefficients can be expanded in a complete set of polynomials in $\theta$ and $\theta'$. This expansion is sufficient and is a complete representation of the $\zeta$ in this case. However, since the $\bar{\varphi}$'s may be expanded in spherical harmonics (\ref{spherical_harmonics_expansion}) and appear in the expressions for the transport coefficients, an expansion of $\zeta$ in terms of the spherical harmonics would be most convenient in computing $\eta, \kappa$. With this objective, an extrapolation of  the two-dimensional function $\zeta(\vec{\bf{q}},\vec{\bf{q}}')$ is made to a function in three dimensions, viz. $\zeta(\vec{\bf{p}},\vec{\bf{p}}')$. This extrapolation is not unique and it is possible to choose one that is convenient for the purposes of this calculation. Since $\vec{\bf{p}}$ and $\vec{\bf{p}}\,'$ are independent vectors, $\theta$ and $\theta'$ dependencies are independent and we can hence assume, separable. Hence, consider the particular expansion
\begin{equation}
\zeta(\theta,\theta',\phi,\phi') = \Sigma_{lm} \zeta_{lm}\,P_{lm}(\theta)\,P_{lm}(\theta')\,e^{im\phi}\,e^{-im\phi'}\,\,\,.
\end{equation}
Here, $P_{lm}$ are the associated Legendre polynomials.
The particular choice of the same indices $(l,m)$ on $P_{lm}(\theta)$ and $P_{l'=l;m'=m}(\theta')$ is one of convenience. While all choices $l'\neq l;m'\neq m$ would yield the correct physical projection $\zeta$ in two-dimensions, the choice $l'=l$ is one of convenience. $m=m'$ is dictated by the isotropy of the function in two-dimensions viz., $\zeta(\phi,\phi') = \zeta(\phi-\phi')$.
This expansion is a more general, yet complete, expansion for $\zeta$ in three dimensions and the former expansion is a projection of this one. Consider a more general $\zeta$ of the form above, or equivalently,
\begin{equation}
\label{zeta_expansion}
\zeta(\theta,\theta',\phi,\phi') = \Sigma_{lm} \zeta_{lm}\,Y_{lm}^{\star}(\theta,\phi)\,Y_{lm}(\theta',\phi')\,\,\,.
\end{equation}
Here, $Y_{lm}$ are the spherical harmonics (Appendix D).
The calculations simplify by generalizing $\zeta$ in this fashion. Clearly, the solution certainly includes the particular $\zeta$ of the rough surface in question and is the correct solution for it. Moreover, the $P_{lm}$ are a set of complete orthogonal polynomials and with $\theta$ and $\theta'$ dependences being separable, offer a correct expansion of $\zeta_m(\theta,\theta')$.

\begin{center}
{\section{Calculation of Viscosity}}
\end{center}
\noindent
The truncated expansion for $\bar{\varphi}$ is
\begin{equation}
\label{phi21expansion}
\bar{\varphi}(\theta,\phi) = \bar{\varphi}_{2,1}\,Y_{2,1}(\theta,\phi) + \bar{\varphi}_{2,-1}\,Y_{2,-1}(\theta,\phi)\,\,\,.
\end{equation}
Projecting the $Y_{21}^{\star}(\theta,\phi)$ component of equation(\ref{eta2}), and using the  properties (Appendix C) of the spherical harmonics, 
\begin{equation}
\label{phi21}
\bar{\varphi}_{2,1} = \left\{\sqrt{30}\left(\zeta_{41}\,B_{41}\,A_{21} + \zeta_{21}\,D_{21}^2 - A_{21}\,B_{41}\,\zeta_{20} - D_{21}^2\,\zeta_{20} - \frac{D_{21}}{3}\,(\zeta_{00}-\zeta_{20})\right)\right\}^{-1}\,\,\,.
\end{equation}
Along the same lines, we get by projecting the $Y_{2,-1}^{\star}(\theta,\phi)$ component of equation(\ref{eta2}),
\begin{equation}
\label{phi2-1}
\bar{\varphi}_{2,-1}=\left\{\sqrt{30}\left(A_{2-1}\,B_{4-1}\,\zeta_{20} + D_{2-1}^2\,\zeta_{20} + \frac{(\zeta_{00}-\zeta_{20})}{3}\,D_{2-1} - B_{4-1}\,\zeta_{4-1}\,A_{2-1} - \zeta_{2-1}\,D_{2-1}^2\right)\right\}^{-1}\,\,\,.
\end{equation}
Putting $\bar{\varphi}_{2\pm1}$ in equation(\ref{eta3}), we get equation(\ref{eta4}) for the viscosity, with
\begin{eqnarray}
\varpi &=& \left(\zeta_{41}\,B_{41}\,A_{21} + \zeta_{21}\,D_{21}^2 - A_{21}\,B_{41}\,\zeta_{20} - D_{21}^2\,\zeta_{20} - \frac{D_{21}}{3}\,(\zeta_{00}-\zeta_{20})\right)^{-1}
\\
&\,\,\,& \,\,\,\,\,\,\,+ \left(-A_{2-1}\,B_{4-1}\,\zeta_{20} - D_{2-1}^2\,\zeta_{20} - \frac{(\zeta_{00}-\zeta_{20})}{3}\,D_{2-1} + B_{4-1}\,\zeta_{4-1}\,A_{2-1} + \zeta_{2-1}\,D_{2-1}^2\right)^{-1}\,\,\,.
\nonumber
\end{eqnarray}

\begin{center}
\section{Calculation of Thermal Conductivity}
\end{center}

\noindent
 The truncated  expansion for $\bar{\varphi}$ is
\begin{equation}
\label{phik_expansion}
\bar{\varphi}(\theta,\phi) = \bar{\varphi}_{1,1}\,Y_{1,1}(\theta,\phi) + \bar{\varphi}_{1,-1}\,Y_{1,-1}(\theta,\phi)\,\,\,.
\end{equation}
Projecting out $Y_{1\pm}^{\star}$  components of equation(\ref{phi_kappa}), we get
\begin{eqnarray}
\label{phi11}
\bar{\varphi}_{1,1} &=& -\left(\sqrt{6}(\zeta_{20}\,A_{11}\,B_{31} + D_{11}^2\,\zeta_{20} + \frac{(\zeta_{00}-\zeta_{20})}{3}\,D_{11} - A_{11}\,B_{3-1}\,\zeta_{3-1} - D_{11}\,D_{1-1}\,\zeta_{1-1})\right)^{-1}\,\,\,;
\nonumber
\\
\bar{\varphi}_{1,-1} &=& \left(\sqrt{6}(\zeta_{20}\,A_{1-1}\,B_{3-1} + D_{1-1}^2\,\zeta_{20} + \frac{(\zeta_{00}-\zeta_{20})}{3}\,D_{1-1} - A_{1-1}\,\zeta_{31}\,B_{3-1} - D_{1-1}\,D_{11}\,\zeta_{11})\right)^{-1}\,\,\,.
\nonumber
\\
&\,&\,\,\,\,\,\,\,\,\,\,\,\,\,\,\,\,\,\,\,\,
\end{eqnarray}
Putting these is equation(\ref{kxx2}), we get equation(\ref{kxx3}) for $\kappa_{xx}$ with
\begin{eqnarray}
\varpi_{\kappa} &=& \Big((\zeta_{20}\,A_{1-1}\,B_{3-1} + D_{1-1}^2\,\zeta_{20} + \frac{(\zeta_{00}-\zeta_{20})}{3}\,D_{1-1} - A_{1-1}\,\zeta_{31}\,B_{3-1} - D_{1-1}\,D_{11}\,\zeta_{11})^{-1}
\nonumber
\\
&+& (\zeta_{20}\,A_{11}\,B_{31} + D_{11}^2\,\zeta_{20} + \frac{(\zeta_{00}-\zeta_{20})}{3}\,D_{11} - A_{11}\,\zeta_{31}\,B_{3-1} - D_{11}\,D_{1-1}\,\zeta_{11})^{-1}\Big)
\end{eqnarray}

\begin{center}
{\section{Properties of Spherical Harmonics}}
\end{center}
\noindent
We use spherical harmonics defined as
\begin{equation}
\label{YP}
Y_{lm}(\theta,\phi) = \sqrt{(2l+1)\frac{(l-m)!}{(l+m)!}}\,P_l^m(cos\theta)\,e^{im\phi}
\end{equation}
with normalization thus
\begin{equation}
\label{Y_orthogonality}
\int\nolimits\frac{d\Omega}{4\pi}\,Y_{lm}(\Omega)\,Y_{l'm'}^{\star}(\Omega) = \delta_{ll'}\,\delta_{mm'}\,\,\,.
\end{equation}
where $P_l^m$ is the associated Legendre polynomial. Now, as worked out in \cite{AppMathLett}, there is a relation
\begin{equation}
\label{appmathlett}
z\,P_n^m(z) = \frac{1}{(2n+1)}\left((n-m+1)\,P_{n+1}^m(z) + (n+m)\,P_{n-1}^m(z)\right)\,\,\,.
\end{equation}
Using equation(\ref{appmathlett}) recursively, 
\begin{equation}
\label{appmathlett2}
z^2\,P_n^m(z)\,e^{im\phi} = \mathcal{C}_{nm}\,P_{n+2}^m(z)\,e^{im\phi} + \mathcal{C'}_{nm}\,P_{n-2}^m(z)\,e^{im\phi} + \mathcal{C''}_{nm}\,P_n^m(z)\,e^{im\phi}\,\,\,,
\end{equation}
where the constants $\mathcal{C}$s are defined thus :
\begin{eqnarray}
\mathcal{C}_{nm} &=& \frac{(n-m+1)(n-m+2)}{(2n+1)(2n+3)}
\nonumber
\\
\mathcal{C'}_{nm} &=& \frac{(n+m)(n+m-1)}{(2n+1)(2n-1)}
\nonumber
\\
\mathcal{C''}_{nm} &=& \frac{(n-m+1)(n+m+1)}{(2n+1)(2n+3)} + \frac{(n+m)(n-m)}{(2n+1)(2n-1)}\,\,\,.
\end{eqnarray}
Define
\begin{equation}
C = \sqrt{(2l+1)\frac{(l-m)!}{(l+m)!}}
\end{equation}
and with equation(\ref{YP}) and $z\equiv cos\theta$, equation(\ref{appmathlett2}) becomes
\begin{eqnarray}
z^2\,Y_{nm}(\theta,\phi) = A_{nm}\,Y_{n+2,m}(\theta,\phi) + B_{nm}\,Y_{n-2,m}(\theta,\phi) + D_{nm}\,Y_{nm}(\theta,\phi)
\nonumber
\\
with\,\,\, A_{nm} = \frac{\mathcal{C}_{nm}}{C_{n+2,m}}\,\,\,;\,\,\,B_{nm} = \frac{\mathcal{C'}_{nm}}{C_{n-2,m}}\,\,\,;\,\,\,D_{nm} = \frac{\mathcal{C''}_{nm}}{C_{nm}}\,\,\,.
\end{eqnarray}

\pagebreak

\pagebreak
\begin{figure}
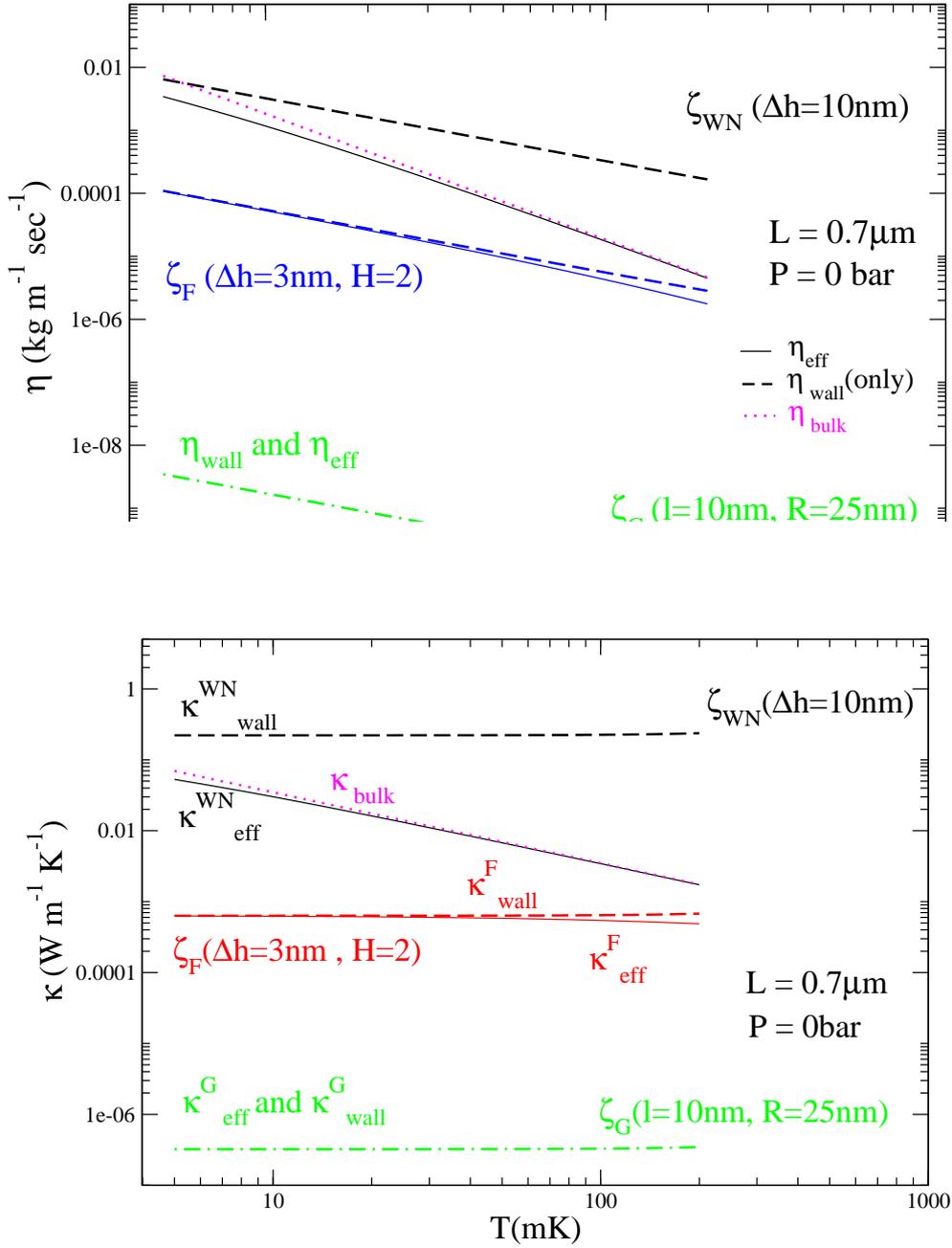

\begin{center}
\includegraphics[%
  width=0.78\linewidth,
  keepaspectratio]{EtavsTloglogp0.eps}
  \includegraphics[%
  width=0.75\linewidth,
  keepaspectratio]{KappavsTLogLogp0.eps}
\end{center}
\caption{Top panel : Calculated values of viscosity(top) and thermal conductivity(bottom) for a film of thickness $L=700$nm and at pressure $P=0$bar. The dotted line shows the bulk value, $\eta_b \propto T^{-2}$ or respectively, $\kappa_b \propto T^{-1}$. Bulk values, including $\tau_0(P)$ are derived from \cite{WheatleyRevModPhys}. The dashed lines show the calculated values with the "mixed" scattering channel. The solid lines show the effective values including both bulk and calculated contributions.  The saturation at low temperatures to the wall-induced component is clear in the fractal case, where we have used rms roughness $\Delta h=3$nm and fractal exponent $H=2$. For white noise roughness with rms roughness $\Delta h = 10$nm, the saturation from the bulk to the wall-component is pushed to lower temperatures. For rougher surfaces, the wall-induced term dominates. Gaussian roughness with height $l=10$nm and correlation length $R=25$nm is used in the figure shown.}
\end{figure}

\end{document}